\documentclass[vecphys]{svmult}

\usepackage{makeidx}         
\usepackage{graphicx}        
\usepackage{multicol}        
\usepackage[bottom]{footmisc}

\makeindex             

\begin{document}

\title*{Principal Component Analysis as a tool to explore star formation histories}

\author{Ignacio Ferreras\inst{1,2}, Ben Rogers\inst{1}\and Ofer Lahav\inst{2}}
\institute{Department of Physics, King's College London. Strand, 
London WC2R 2LS, U.K.\texttt{ferreras@star.ucl.ac.uk}
\and Department of Physics and Astronomy, University College London. 
Gower St., London WC1E 6BT, UK}
%
%
\maketitle
\index{Ferreras}
\index{Rogers}
\index{Lahav}

Principal Component Analysis (PCA) is a well-known multivariate
technique used to decorrelate a set of vectors. PCA has been
extensively applied in the past to the classification of stellar and
galaxy spectra (e.g. Madgwick et al. 2002). Here we apply PCA to the
optical spectra of early-type galaxies, with the aim of extracting
information about their star formation history.  We consider two
different data sets: 1) a reduced sample of $30$ elliptical galaxies
in Hickson compact groups and in the field, and 2) a large
volume-limited ($z<0.1$) sample of $\sim 7,000$ galaxies from the
Sloan Digital Sky Survey. Even though these data sets are very
different, the homogeneity of the populations results in a very
similar set of principal components. Furthermore, most of the
information (in the sense of variance) is stored into the first few
components in both samples. The first component (PC1) can be
interpreted as an old population and carries over 99\% of the
variance.  The second component (PC2) is related to young stars and we
find a correlation with NUV flux from GALEX. Model fits consistently
give younger ages for those galaxies with higher values of PC2.

\section{Why PCA on early-type galaxy spectra ?}
\label{sec:1}

Unravelling the star formation history of massive early-type galaxies
is one of the key issues towards a complete theory of galaxy
formation. These systems pose a strong constraint on the standard
paradigm of hierarchical structure formation, as they are massive
enough to have undergone an extended period of assembly, whereas their
light is dominated by old populations, revealing an early, short and
intense star formation history (e.g. Trager et al. 2000). Recent
searches for massive galaxies at high redshift reveal a population
of red and massive systems at $z>2$ whose photometry, mass
distribution and spatial correlation properties make them plausible
candidates for the progenitors of local early-type galaxies (e.g. 
Labb\'e et al. 2003; Daddi et al. 2003; van Dokkum et al 2006).

``Traditional'' methods to assess the stellar populations in galaxies
rely on direct comparisons with models, including a standard maximum
likelihood approach to determine the best fit and the uncertainties on
the parameters that control the models (Panter, Heavens \&
Jimenez 2004; Ocvirk et al. 2006). It has been shown (e.g. Ferreras \& Yi
2004) that large degeneracies exist in the determination of the age of
an old stellar population. Therefore, a model-independent approach
is most welcome in this field, and PCA is one such method.

Principal Component Analysis (PCA; e.g. Rencher 2002) is a multivariate
analysis tool aimed at looking for linear combinations of the original
data so that the new elements (called principal components) are statistically
uncorrelated.  The methodology reduces to a diagonalisation of the covariance
matrix, so that the eigenvectors are the principal
components and the eigenvalues give the weight of each one. Usually
these components are sorted in decreasing order of their eigenvalues,
with the first components carrying most of the information (in the
sense of variance).

\begin{figure}[t]
\centering
\includegraphics[height=7cm,width=9cm]{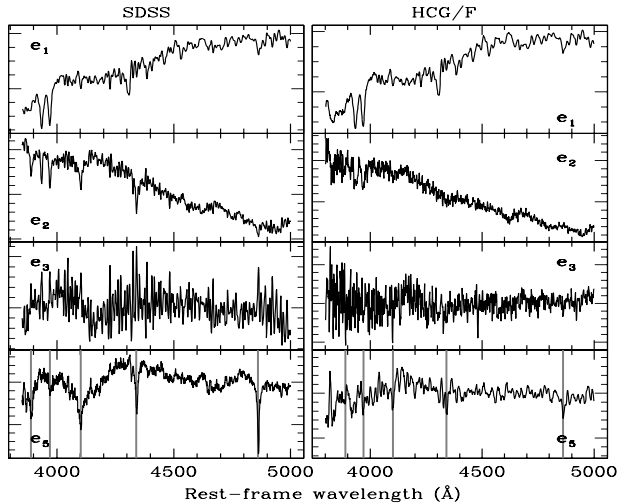}
\caption{Principal components corresponding to the
two samples described in this paper: {\sl left} SDSS early-type
galaxies, {\sl right} Compact Group/Field elliptical galaxy sample.
Notice the similarity of all components, and the presence of the
Balmer series (grey lines) in $e_5$.}
\label{fig:PCs} 
\vskip-0.15in
\end{figure}

\section{The samples}
\label{sec:2}

We have applied PCA to two sets of spectroscopic data from early-type
galaxies.  The first sample comprises 30 elliptical galaxies located
in Compact Groups and in the field (Ferreras et al. 2005), hereafter
referred to as HCG/F. The other sample is a volume-limited set
($z<0.1$) of early-type systems extracted from the Sloan Digital Sky
Survey (Bernardi et al. 2005). In both cases the spectra were acquired
with the same instrument for all galaxies, a crucial issue when
applying PCA. The HCG/F sample comes from observations at the KPNO
2.1m telescope (de la Rosa et al. 2001) whereas the SDSS sample come
from the publicly available Data Release~4 (Adelman-McCarthy et
al. 2006).  In both sets the spectra were de-redshifted and
de-reddened (from Galactic extinction) using standard techniques. The
spectral range 3800--5000\AA\  was used.  The reader is pointed to
Ferreras et al. (2005) for details of the Compact Group/Field sample
and to Rogers et al. (in preparation) for details of the SDSS sample.

\begin{figure}[t]
\centering
\includegraphics[height=7cm,width=9cm]{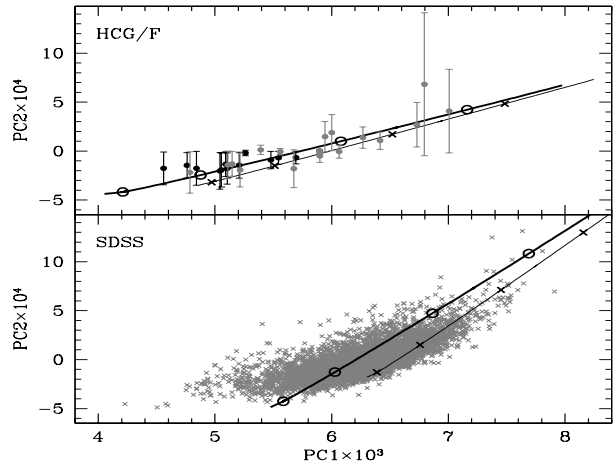}
\caption{Projection of the galaxy spectra on to the first and second
principal components for both samples. The lines represent an
exponentially-decaying star formation history as described in the
text, with symbols from bottom-left at a timescale of $\tau_s =11, 10, 9$
and $8$~Gyr. The thick (thin) line corresponds to a metallicity of
$2.5Z_\odot$ ($Z_\odot$).}
\label{fig:PC12} 
\vskip-0.15in
\end{figure}

\section{The Principal Components}
\label{sec:3}

Treating the fluxes in each wavelength bin ($\{\Phi
(\lambda_i)\}_{i=0}^N$) as the components of a vector in N-dimensional
space, one can construct the covariance matrix for the set of
galaxies.  The first eigenvectors are shown in figure~\ref{fig:PCs}
for both samples. The first and second eigenvectors stand out as
representing old and young stellar populations, respectively. The third
eigenvector is not so easy to interpret.  We find that it is related to
velocity dispersion (Ferreras et al. 2005; Rogers et al. in
preparation). Finally, it is remarkable to find {\sl in both sets}
that the fifth component features the characteristic Balmer absorption
lines typical of A-type stars. This feature is commonly interpreted as
a signature of a recent star formation episode.

In order to test the ``success'' of PCA in classifiying the spectra,
we project each galaxy on to the eigenvectors.  PCA can be thought of
as a ``rotation'' in an N-dimensional space, so that the new axes (the
principal components) are optimally aligned towards the directions of
maximal variance. Given that most of the weight lies on the few first
eigenvectors (in both samples the first three components amount for
more than 99\% of the information), one only needs to explore these
ones. Figure~\ref{fig:PC12} shows the result for the projection on to
the first and second components (PC1 and PC2 respectively). Each point
represents a galaxy. In the HCG/F sample field (compact group)
galaxies are represented as black (grey) circles.  The lines are
projections of synthetic spectra on to the same principal
components. These spectra are composite $\tau$-models from Bruzual \&
Charlot (2003), assuming an exponential star formation rate. Various
values of the timescale ($\tau_s$) are explored. The open circles
(crosses) represent -- from bottom-left -- values of
$\tau_s=11,10,9,8$~Gyr for metallicity $2.5Z_\odot$ ($Z_\odot$).  The
observed correlation between PC1 and PC2 in these galaxies shows
roughly an age trend, so that galaxies in compact groups feature a
wider distribution of ages compared to their counterparts in the
field. The SDSS data shows a similar range of values of PC1
and PC2.

\begin{figure}[t]
\centering
\includegraphics[height=7cm,width=9cm]{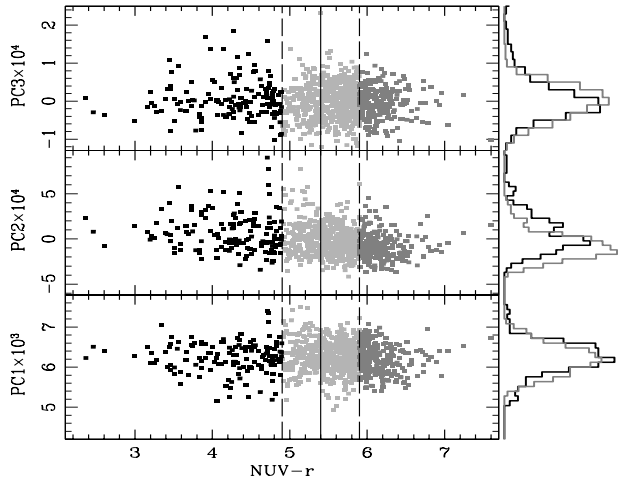}
\caption{Comparison between the projection of the first three
principal components and NUV$-r$, colour, where the NUV is the
GALEX/NUV passband (2300\AA\ ) flux (Schawinski et al. 2006). The histograms on
the right refer to the NUV luminous (black; NUV$-r<4.9$) or the NUV
faint (grey; NUV$-r>5.9$) galaxies. Schawinski et al. (2006) used
NUV$-r<5.4$ as a criterion for recent star formation.
Notice that only PC2 -- which
corresponds to a young component -- correlates with colour.}
\label{fig:galex} 
\vskip-0.15in
\end{figure}

We wanted to explore this point further by cross-matching our SDSS
dataset with the recent sample of Schawinski et al. (2006) who
measured fluxes from GALEX. We use their NUV passband centered around
2300\AA\ . Their criterion for recent star formation was defined on
the NUV$-r$, so that colours bluer than 5.4 are characteristic of
galaxies with young stellar populations.  Figure~\ref{fig:galex} shows
a comparison between NUV$-r$ colour and our projections on to the
first three principal components. Even though a clean correlation is
not apparent, the histograms on the right, obtained for a blue and a
red subsample, show that only PC2 features a correlation with NUV$-r$
and thereby with recent star formation. We should emphasize here that
the estimated mass fraction in young stars stays -- even for the
bluest galaxies in this sample -- below a few percent. Nevertheless, 
this small contribution gives a significant flux in the NUV.  It is
remarkable that our PCA analysis -- restricted to 3800-5000\AA\ -- can
detect these small episodes of star formation~!

\begin{figure}[t]
\centering
\includegraphics[width=5.4cm]{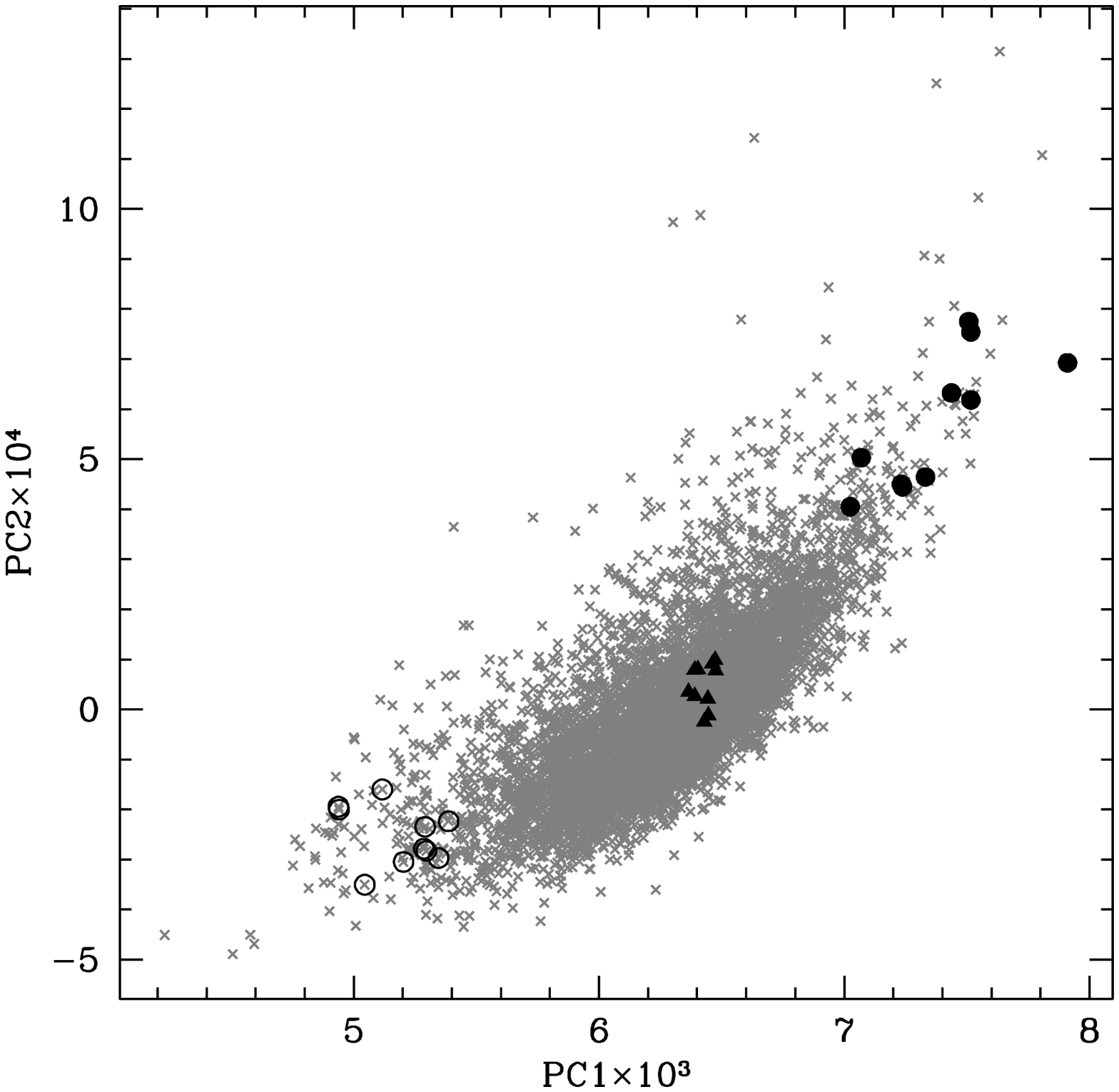}
\includegraphics[width=5.4cm]{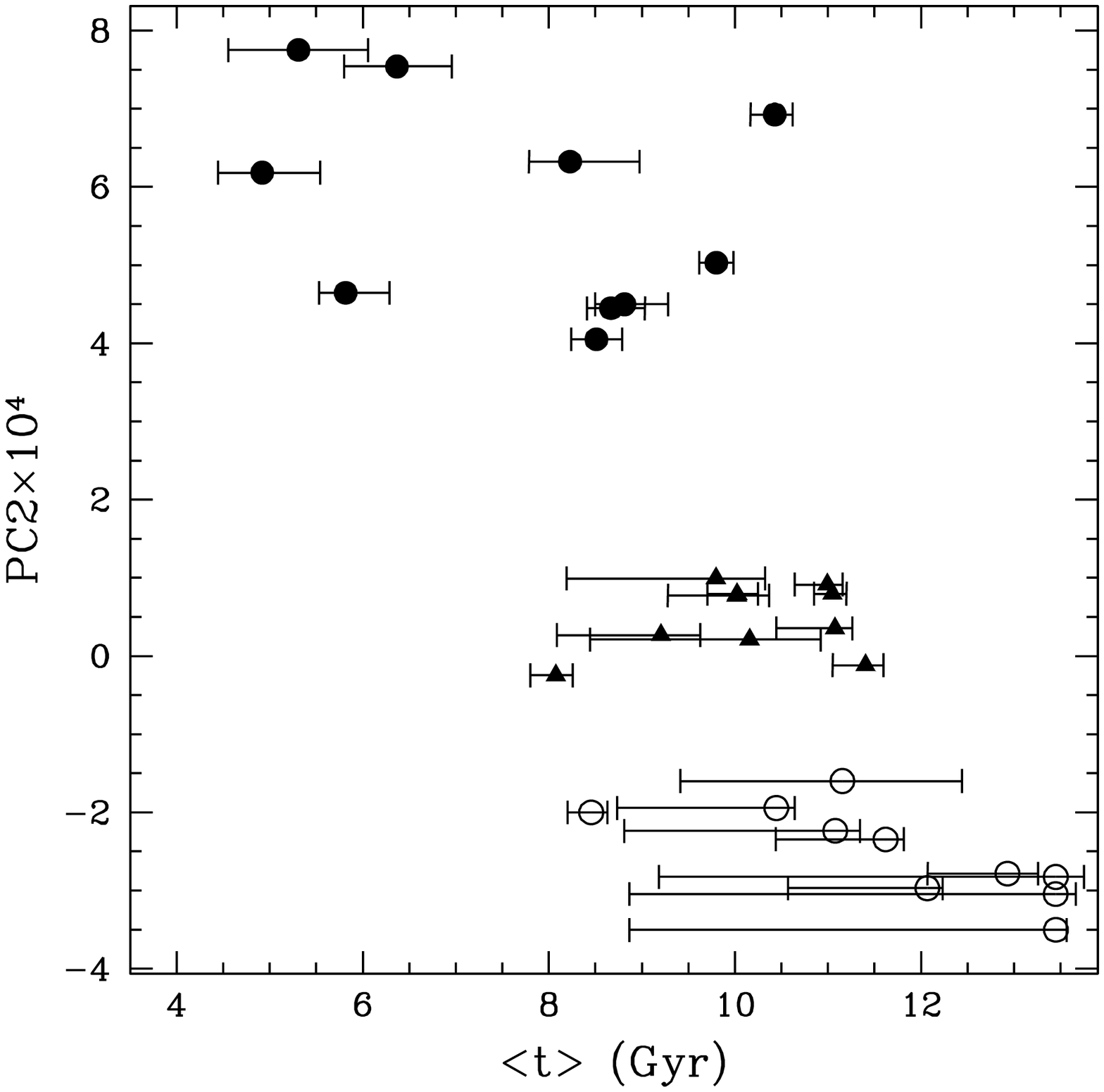}
\caption{A comparison between the projection of the principal
components ({\sl left}) and the estimated ages using 
a ``traditional'' maximum likelihood approach. The models assume an
exponentially decaying star formation history at fixed metallicity. The
error bars are given at the 90\% confidence level. Galaxies from
three regions of the (PC1,PC2) plot are explored, showing that
those with higher values of PC2 indeed show younger ages.}
\label{fig:chi2} 
\vskip-0.15in
\end{figure}

\section{Putting the physics back in}
\label{sec:4}

The weakness of PCA ironically lies in its strength. PCA is a
model-independent method, which only reclassifies input data according
to its covariance. Hence, we cannot give direct physical meaning to
the eigenvectors. Nevertheless, one can put the physics {\sl a
posteriori}, by projecting synthetic models on to the principal
components as shown above, or we can use the classification suggested
by PCA to select specific subsets of galaxies for comparison with
models. The use of PC2 as a proxy for recent star formation is further
explored by selecting a few galaxies from the PC1 vs. PC2 plane (see
left panel of figure~\ref{fig:chi2}).  These galaxies are extracted
from different regions of the diagram. The spectra are compared with
$\tau$-models (exponentially decaying star formation rate at constant
metallicity) via a maximum likelihood method, and the results for the
average age are shown in the right panel of figure~\ref{fig:chi2}. The
error bars represent the 5 and 95 percentile of the probability
distribution. The correlation with PC2 is evident, confirming our
preliminary results. The range of ages is the one expected for massive
early-type galaxies (Trager et al. 2000).

\section{Conclusions}
\label{sec:5}
The work presented here represents a pilot study aimed at exploring
PCA as a complement (never a substitute) to model-dependent techniques
to learn about the star formation history of galaxies from their
unresolved photo-spectroscopic information. We have shown that some of
the principal components have a direct connection with physical
parameters. Most notably PC2 correlates with the distribution of
stellar ages. No space is left in this paper to discuss other
interesting components, such as PC3 or PC5, but we refer the reader to
our ongoing work on the SDSS sample (Rogers et al.  in preparation).

\medskip
\noindent
ACKNOWLEDGEMENT: The work presented here combines two independent
data sets on which PCA was applied separately. This work would not 
have been possible without the help of our collaborators Anna Pasquali,
Reinaldo de~Carvalho, Mariangela Bernardi, Sugata Kaviraj and Sukyoung Yi.
The SDSS collaboration is gratefully acknowledged for frequent use
of its superb data base.




\printindex

\begin{thebibliography}{99.}

\bibitem{DR4} Adelman-McCarthy, et al. ApJS, \textbf{162}, 38 (2006).

\bibitem{ber} M. Bernardi, et al. AJ, \textbf{129}, 61 (2005).

\bibitem{BC03} G. Bruzual \& S. Charlot, MNRAS, \textbf{344}, 1000 (2003).

\bibitem{dad03} E. Daddi, et al., ApJ, \textbf{588}, 50 (2003).

\bibitem{dlr01} I. G. de la Rosa, R.~R. de~Carvalho \& S.~E. Zepf, AJ, \textbf{122}, 93 (2001).

\bibitem{HCG} I. Ferreras, A. Pasquali, R.~R. de~Carvalho, I.~G. de~la~Rosa
\& Lahav, O., MNRAS, \textbf{123}, 456 (2005).

\bibitem{lbds} I. Ferreras \& S. Yi, MNRAS \textbf{350}, 1322 (2004).

\bibitem{lab03} I. Labb\'e, et al. AJ, \textbf{125}, 1107 (2003).

\bibitem{madg} D.~S. Madgwick, et al. MNRAS, \textbf{333}, 133 (2002).

\bibitem{MAP} P. Ocvirk, et al. MNRAS, \textbf{365}, 46O (2006).

\bibitem{moped} B. Panter, A.~F. Heavens\ \& R. Jimenez, MNRAS, \textbf{343}, 1145 (2003).

\bibitem{MVA} A.~C. Rencher: \textit{Methods of Multivariate Analysis}, 2nd edn 
(Wiley, New York 2002) pp 380--407.

\bibitem{Galex} K. Schawinski, et al., astro-ph/0601036 (2006).

\bibitem{sct} S.~C. Trager, et al. AJ, \textbf{120}, 165 (2000).

\bibitem{vdk06} P. G. van Dokkum, et al., ApJ, \textbf{638}, L59 (2006).

\end{thebibliography}
\end{document}